\documentclass[conference]{IEEEtran}
\ifCLASSINFOpdf
  \usepackage[pdftex]{graphicx}
\else
  \usepackage[dvips]{graphicx}
\fi
\usepackage[cmex10]{amsmath}
\usepackage{amssymb}
\usepackage{mathtools}
\usepackage{array}

\usepackage{algpseudocode,algorithm,algorithmicx}
\usepackage{float}
\usepackage{tabularx}
\usepackage{mathrsfs} 
\usepackage{psfrag}
\usepackage{tikz}
\usetikzlibrary{chains, arrows,shapes,positioning,arrows,decorations.markings,calc, spy}
\usepackage{pgfplots}
\pgfplotsset{compat=newest}
\usepgfplotslibrary{groupplots}
\usepgfplotslibrary{polar}
\usepackage{gensymb}
\usepackage{xcolor}
\usepackage{todonotes}
\usepackage[shortcuts, acronym]{glossaries}
\usepackage[caption=false]{subfig}
\DeclareMathOperator*{\argmin}{argmin}

\pgfplotsset{
    mark repeat*/.style={
        scatter,
        scatter src=x,
        scatter/@pre marker code/.code={
            \pgfmathtruncatemacro\usemark{
                or(mod(\coordindex,#1)==0, (\coordindex==(\numcoords-1))
            }
            \ifnum\usemark=0
                \pgfplotsset{mark=none}
            \fi
        },
        scatter/@post marker code/.code={}
    }
}

\usepackage{fixmath}
\usepackage{amssymb}
\usepackage{bm}
\renewcommand{\arraystretch}{2}
\usepackage{pbox}
\usepackage{algorithm}
\usepackage{lettrine}

\algnewcommand{\IIf}[1]{\State\algorithmicif\ #1\ \algorithmicthen}
\algnewcommand{\EndIIf}{\unskip\ \algorithmicend\ \algorithmicif}

\definecolor{BTORMIX1}{rgb}{0.125, 0,0.875}
\definecolor{BTORMIX2}{rgb}{0.25, 0,0.75}
\definecolor{BTORMIX3}{rgb}{0.375,0,0.625}
\definecolor{BTORMIX4}{rgb}{0.50,0,0.50}
\definecolor{BTORMIX5}{rgb}{0.625,0,0.375}
\definecolor{BTORMIX6}{rgb}{0.75,0,0.25}
\definecolor{BTORMIX7}{rgb}{0.875,0,0.125}

\definecolor{BTOGMIX1}{rgb}{0, 0.125, 0.875}
\definecolor{BTOGMIX2}{rgb}{0, 0.25, 0.75}
\definecolor{BTOGMIX3}{rgb}{0, 0.375, 0.625}
\definecolor{BTOGMIX4}{rgb}{0, 0.50, 0.50}
\definecolor{BTOGMIX5}{rgb}{0, 0.625, 0.375}
\definecolor{BTOGMIX6}{rgb}{0, 0.75 ,0.25}
\definecolor{BTOGMIX7}{rgb}{0, 0.875, 0.125}

\begin{document}
%
\title{Robust massive MIMO Equilization for mmWave systems with low resolution ADCs}
\author{\IEEEauthorblockN{Kilian Roth \IEEEauthorrefmark{1}\IEEEauthorrefmark{2}, Josef A. Nossek\IEEEauthorrefmark{2}\IEEEauthorrefmark{3}}
\IEEEauthorblockA{\IEEEauthorrefmark{1} Next Generation and Standards, Intel Deutschland GmbH, Neubiberg, Germany \\
Email: $\{$kilian.roth$\}$@intel.com}
\IEEEauthorblockA{\IEEEauthorrefmark{2} Department of Electrical and Computer Engineering,
 Technical University Munich, Munich, Germany \\
Email: $\{$kilian.roth, josef.a.nossek$\}$@tum.de}
\IEEEauthorblockA{\IEEEauthorrefmark{3} Department of Teleinformatics Engineering, Federal University of Ceara, Fortaleza, Brazil}}
\maketitle
\newacronym{A/D}{A/D}{Analog/Digital}
\newacronym[plural=ADCs,firstplural=Analog-to-Digital-Converters (ADCs)]{ADC}{ADC}{Analog-to-Digital-Converter}
\newacronym{AGC}{AGC}{Automatic Gain Control}
\newacronym{AQNM}{AQNM}{Additive Quantization Noise Model}
\newacronym{BB}{BB}{BaseBand}
\newacronym{BER}{BER}{Bit Error Rate}
\newacronym{BLER}{BLER}{BLock Error Rate}
\newacronym[plural=DACs,firstplural=Digital-to-Analog-Converters (ADCs)]{DAC}{DAC}{Digital-to-Analog-Converter}
\newacronym{CDM}{CDM}{Code Division Multiplex}
\newacronym{CDMA}{CDMA}{Code Division Multiple Access}
\newacronym{CIR}{CIR}{Channel Impulse Response}
\newacronym{CMOS}{CMOS}{Complementary Metal–Oxide–Semiconductor}
\newacronym{CS}{CS}{Cyclic Shift}
\newacronym{CP}{CP}{Cyclic Prefix}
\newacronym{DBF}{DBF}{Digital BeamForming}
\newacronym{DFT-s-OFDM}{DFT-s-OFDM}{DFT-spread-OFDM}
\newacronym{DMRS}{DMRS}{DeModulation Reference Signals}
\newacronym{DCD}{DCD}{Dichotomous Coordinate Descent}
\newacronym{EVM}{EVM}{Error Vector Magnitude}
\newacronym{EM}{EM}{Expectation Maximization}
\newacronym{FDM}{FDM}{Frequency Division Multiplex}
\newacronym{HBF}{HBF}{Hybrid BeamForming}
\newacronym{ISM}{ISM}{Industrial, Scientific and Medical}
\newacronym{LA}{LA}{Limiting Amplifier}
\newacronym[plural=LOs,firstplural=Local Oscillators (LOs)]{LO}{LO}{Local Oscillators}
\newacronym{LTE}{LTE}{Long Term Evolution}
\newacronym{MIMO}{MIMO}{Multiple Input Multiple Output}
\newacronym{MINLP}{MINLP}{Mixed Integer Non-Linear Programing}
\newacronym{mmWave}{mmWave}{millimeter Wave}
\newacronym{MMSE}{MMSE}{Minimum Mean Square Error}
\newacronym{MSE}{MSE}{Mean Square Error}
\newacronym{ML}{ML}{Most Likelyhood}
\newacronym{MU-MIMO}{MU-MIMO}{Multi User - Multiple Input Multiple Output}
\newacronym{NLP}{NLP}{NonLinear Programing}
\newacronym{NR}{NR}{New Radio}
\newacronym{OFDM}{OFDM}{Orthogonal Frequency Domain Multiplexing}
\newacronym{PDP}{PDP}{Power Delay Profile}
\newacronym[plural=PAs,firstplural=Power Amplifiers (PAs)]{PA}{PA}{Power Amplifier}
\newacronym{PN}{PN}{Pseudo Noise}
\newacronym{QAM}{QAM}{Quadrature Amplitude Modulation}
\newacronym{Q}{Q}{Quantization}
\newacronym{RE}{RE}{Resource Element}
\newacronym{RF}{RF}{Radio Frequency}
\newacronym{RFE}{RFE}{Radio Front-End}
\newacronym{RMS}{RMS}{Root Mean Square}
\newacronym{Rx}{Rx}{receiver}
\newacronym[plural=SISOs,firstplural=Single Input Single Output (SISOs)]{SISO}{SISO}{Single Input Single Output}
\newacronym{SNR}{SNR}{Signal to Noise Ratio}
\newacronym[plural=SCs,firstplural=Sub-Carriers (SCs)]{SC}{SC}{Sub-Carrier}
\newacronym{Tx}{Tx}{transmitter}
\newacronym[plural=UE,firstplural=User Equipment (UE)]{UE}{UE}{User Equipment}
\newacronym{ULA}{ULA}{Uniform Linear Array}
\newacronym{UL}{UL}{Uplink}
\newacronym{VGA}{VGA}{Variable Gain Amplifier}

\begin{abstract}
Leveraging the available millimeter wave spectrum will be important for 5G.
In this work, we investigate the performance of digital beamforming with low resolution ADCs based on link level simulations including channel estimation, MIMO equalization and channel decoding. 
We consider the recently agreed 3GPP NR type 1 OFDM reference signals. The comparison shows sequential DCD outperforms MMSE-based MIMO equalization both in terms of
detection performance and complexity. We also show that the DCD based algorithm is more robust to channel estimation errors. 
In contrast to the common believe we also show that the complexity of MMSE equalization for a massive MIMO system is not dominated by the matrix inversion but by the computation of the Gram matrix. 
\end{abstract}
%
%
%
%
%

\section{Introduction}
For future \ac*{mmWave} mobile broadband systems, analog/hybrid beamforming are considered to be a possible solution to the excessive power consumption at the receiver. Due to the large bandwidth, high resolution \acp*{ADC} have a significant amount of power. Therefore, they are considered to be a major contributor to the power consumption of a \ac*{mmWave} receiver. 

Analog/hybrid beamforming highly depend on the optimal alignment of beams. The required beam-training/alignment has to be implemented as a search procedure, essentially different configuration are tested and the best one is selected \cite{WIGIGSTDORIGINAL}. Considering such a procedure for multiple UEs at the same time can be considered to have a large overhead. A possible solution to these type of systems is digital beamforming with low resolution \acp*{ADC}. As we showed in \cite{KILIANJSACEE} the power consumption of the receiver frontend of both systems is comparable. In most cases the low resolution \ac*{ADC} digital beamforming is most energy efficient. 
A lot of the work on low resolution \ac*{ADC} digital beamforming only considers the extreme case of 1-bit quantization for the inphase and quadrature \cite{Mo2014, Choi2016, Li2017}.
As we showed in \cite{KILIANJSACEE} it is not clear that 1-bit quantization does lead to the most energy efficient implementation, we consider low resolution \acp*{ADC} in general. 

Many of the investigated channel estimation and detection schemes require  detailed knowledge about the structure of the channel. For example algorithms like GAMP \cite{Mo2014, Choi2016} are very sensitive to the case that their modeling assumptions are not fully valid. 
 Other algorithms like \ac*{EM} require accurate knowledge about the sparsity or related parameters \cite{Mo2014, Choi2016}. In a practical system, this knowledge is hard to obtain. In addition, the systems should robust regarding cases, where the assumptions leading to a specific algorithm are not fully satisfied.

Many massive \ac*{MIMO} equalization schemes consider only perfect channel estimation \cite{Gao2014, WU2016} even without channel coding. We think the propagation of the channel estimation error inside the \ac*{MU-MIMO} equalization is not straight forward. Only for linear methods the influence of the channel estimation error can be investigated theoretically \cite{Eraslan2013}.
This motivated us to investigate channel estimation, \ac*{MIMO} equalization in combination with channel coding and low resolution \acp*{ADC}. 

Therefore, we wanted to investigate the performance of these systems without limiting assumptions on the statistics of the channel. 
We also wanted to put the focus on linear, low complexity algorithms while considering reference signals developed for NR in 3GPP.
Even so we do not consider finite resolution calculation in this work, it is important to mention that massive \ac*{MIMO} is very robust to these effects \cite{Gunnarsson2017}. 
This could further simplify the required calculation and lead to an implementation that might even be feasible for mobile devices. 

In the following sections, we introduce the used channel estimation scheme including a description of the recently in 3GPP agreed NR Type 1 OFDM reference signals. 
Afterwards, we introduce the sequential \ac*{DCD} algorithm for \ac*{MU-MIMO} equalization. In the end, we show a performance and computational complexity based comparison to \ac*{MMSE} \ac*{MU-MIMO} detection.

\section{System Model}

\subsection{NR reference structure}
\begin{figure}
	\centering
	\input{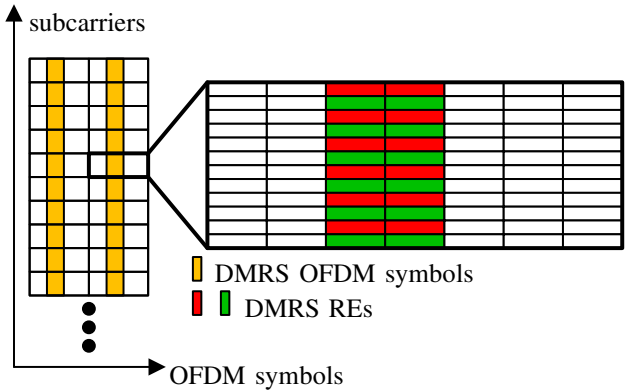}
	\caption{3GPP NR OFDM type 1 DMRS for up to 8 users.}
	\label{fig:NRREFStructure}
\end{figure}

As described in \cite{3gpp.38.211}, the \ac*{OFDM} type 1 \ac*{DMRS} consists of a \ac*{PN} sequence defined by parts of a length-31 Gold sequence.
There are three techniques used to orthogonalize the reference signal for different users:
\begin{itemize}
	\item Time domain \ac*{CS}
	\item \ac*{FDM}
	\item \ac*{CDM} for the case of more than 4 users
\end{itemize}
The resource grid in Fig. \ref{fig:NRREFStructure} shows a possible allocation of the reference symbol \acp*{RE} within a slot.
Among the three ways to orthogonalize the reference signals send by different users or for different MIMO streams, \ac*{FDM} and \ac*{CDM} are most commonly used in many communication systems.
\ac*{CS} is not encountered in many current wireless communication standards except for \ac*{LTE} \ac*{UL}, thus more detailed explanation on how it could provide sufficient orthogonality is needed. 

Since a  \ac*{CS} of half of the \ac*{OFDM} symbol offers the highest distance between two reference sequences used for different users, only this value is foreseen in the standard. 
Due to the properties of the Fourier transformation, a cyclic shift of $N/2$ in the time domain corresponds to a modulation of the frequency domain sequence with the 
sequence $\boldsymbol{c} = [-1, 1, -1, 1, \cdots]^T$. 
The reference symbol sequence $\boldsymbol{s}$ is generated as QPSK symbols with the bits defined by a \ac*{PN} sequence. 
In this case one user sends the sequence $\boldsymbol{s}$ and the other the time domain cyclic shifted version $\boldsymbol{s} \odot \boldsymbol{c}$. 
Assuming perfect synchronization and the delay spread of the channel being smaller than the \ac*{CP}, the linear convolution with the channel is converted to a circular one. 

To demodulate the \ac*{DMRS}, we first transform the signal into the frequency domain. 
Afterwards, we multiply with the complex conjugate of the of the used reference sequence. 
Since this would correspond to a cyclic convolution in the time domain this does still include all possible cyclic shifts and therefore also the one from the cyclic shifted sequence. Thus, additionally we need to apply a windowing in the time domain to limit the interference among the original sequence and its cyclic shift. 
The windowing in the time corresponds to a cyclic convolution in the frequency domain. 
It will be implemented by the spatial smoothing filters shown in the following channel subsection. Combining this observations with the fact that $\boldsymbol{s}$ should be designed to have a cyclic auto-correlation function with only one strong peak and only small values otherwise, we can conclude that a cyclic shift can sufficiently orthogonalize the signal from different users. 

In general we can describe the reference signal of $i$th \ac*{MIMO} layer $r^i$ on sub-carrier $k$ and \ac*{OFDM} symbol $\ell$ as
\begin{equation}
	a^i_{(k, \ell)} = \alpha_{\text{CS}}(i, k) \alpha_{\text{FDM}}(i, k) \alpha_{\text{CDM}}(i, \ell)  [\boldsymbol{s}]_{\lfloor k / 2 \rfloor} ,
\end{equation} 
where $\alpha_{\text{CS}}(i, k)$, $\alpha_{\text{FDM}}(i, k)$ and $\alpha_{\text{CDM}}(i, \ell)$ are the changes of sequence based on \ac*{CDM}, \ac*{FDM} and \ac*{CS}. 
These parameters are defined in the following way
\begin{equation}
\begin{gathered}
	\alpha_{\text{CS}}(i, k) = w^i_{\text{CS}}(\lfloor k / 2 \rfloor \bmod 2), \\
	\alpha_{\text{FDM}}(i, k) = \left\{\begin{array}{ll} 1, & k \bmod 2 = w^i_{\text{FDM}} \\  0, &  \text{otherwise}\end{array}\right. ,\\
	\alpha_{\text{CDM}}(i, \ell) = \left\{\begin{array}{ll} w^i_{\text{CDM}}(0), & \ell = \ell_0 \\  w^i_{\text{CDM}}(1), & \ell = \ell_0 + 1\end{array}\right. ,
\end{gathered}
\end{equation}
where $\ell_0$ is the first \ac*{DMRS} symbol with \ac*{DMRS} in a slot. 
A table of the parameters $w^i_{\text{CS}}$, $w^i_{\text{FDM}}$ and $w^i_{\text{CDM}}$ dependent on the \ac*{MIMO} layer $i$ can be found in Table \ref{tab:NROFDMType1Config}. 
It is important to mention that for the case of 1 to 4 \ac*{MIMO} layers, no \ac*{CDM} based orthogonalization is necessary, 
\begin{table}
	\renewcommand{\arraystretch}{1.3}
	\caption{Configuration of NR type 1 OFDM reference signals. }
	\label{tab:NROFDMType1Config}
	\centering
		\begin{tabular}{|p{1.61cm}|p{2.11cm}|p{0.81cm}|p{2.51cm}|}
			\hline
			MIMO layer $i$ & $[w^i_{\text{CS}}(0), ~ w^i_{\text{CS}}(1)]$ & $w^i_{\text{FDM}}$ & $[w^i_{\text{CDM}}(0),~ w^i_{\text{CDM}}(1)]$ \\ \hline \hline
			$1$ & $[1, ~~1]$ & $0$ & $[1, ~~1]$ \\  \hline
			$2$ & $[1, -1]$ & $0$ & $[1, ~~1]$ \\  \hline
			$3$ & $[1, ~~1]$ & $1$ & $[1, ~~1]$ \\  \hline
			$4$ & $[1, -1]$ & $1$ & $[1, ~~1]$ \\  \hline
			$5$ & $[1, ~~1]$ & $0$ & $[1, -1]$ \\  \hline
			$6$ & $[1, -1]$ & $0$ & $[1, -1]$ \\  \hline
			$7$ & $[1, ~~1]$ & $1$ & $[1, -1]$ \\  \hline
			$8$ & $[1, -1]$ & $1$ & $[1, -1]$ \\  \hline
		\end{tabular}
\end{table}

\subsection{Channel Estimation}
Assuming perfect synchronization of the timing and carrier frequency, the \ac*{OFDM} receive signal $Y_{k, \ell}$ of subcarrier $k$ and \ac*{OFDM} symbol $\ell$ can be written as
\begin{equation}
	Y_{k, \ell,} = H_{k, \ell} X_{k, \ell} + \eta_{k, \ell},
\end{equation}
where we assume that the \ac*{CIR} is shorter than the \ac*{CP}, and $H_{k, \ell}$, $X_{k, \ell}$ and $\eta_{k, \ell}$ are the channel,
 transmit signal and white Gaussian noise of the system, respectively. Time-frequency filters are used to interpolate the channel 
estimates between the position of the reference symbols. A two times 1-D time-frequency interpolation method based on a \ac*{MMSE} criteria as
described in \cite{CEMMSE} is identified as the solution with the best performance complexity trade-off.
First we use a 1-D time-frequency-space filter for smoothing of the estimate on all subcarriers. 
Afterwards, we interpolate and extrapolate the channel on all \ac*{OFDM} symbols.
This procedure is executed for each antenna separately.
It is important to note that this technique assumes knowledge of the following statistical channel parameters:
\begin{itemize}
	\item Doppler spread
	\item Delay spread
	\item Receive \ac*{SNR} of each user
\end{itemize}
As we consider a \ac*{MU-MIMO} scenario, we need to ensure that different users have orthogonal reference sequences. In particular, we will assume that the training sequences are orthogonal.
As we showed in the previous subsection, this can be ensured by the chosen design.
Therefore, the following calculation is done for each user, and thus no user index is included to simplify the notation.

Assuming a reference symbol is present on subcarrier $q$ and symbol time $p$, we multiply the signal with the known reference signal to obtain the corresponding channel estimate for this symbol
\begin{equation}
	\hat{H}_{p, q} = Y_{p, q}  X^*_{p, q} = H_{p, q} + \eta_{p, q},
\end{equation}
where we assume that $\left\vert X^*_{p, q}\right\vert = 1$. 
By combining the channel estimates for all resource elements on $K$ subcarriers and $L$ symbols we get
\begin{equation}
	\hat{\boldsymbol{h}}_r = \left[ \hat{H}_{1, 1}, \hat{H}_{2, 1}, \cdots, \hat{H}_{K-1, L},  \hat{H}_{K, L} \right]^T.
\end{equation}
For all positions where no reference signals were sent, the corresponding element of $\hat{\boldsymbol{h}}_r$ is set to zero.
The set $\mathbb{P}$ contains the indices of the reference symbols in $\hat{\boldsymbol{h}}_r$.

Applying the matrices for interpolation and smoothing in time $\boldsymbol{A}_t$ and frequency $\boldsymbol{A}_f$ domain,
we get the overall estimate of the channel at each position 
\begin{equation}
	\hat{\boldsymbol{h}} = \boldsymbol{A}_{tf} \hat{\boldsymbol{h}}_r = \left(\boldsymbol{A}_t \otimes \boldsymbol{A}_f\right) \hat{\boldsymbol{h}}_r.
\end{equation}
We choose these interpolation matrices separately for each dimension separately to reduce the complexity.
In general to achieve the theoretical optimal performance, these interpolation matrices have to be chosen according
to the covariance matrix of the channel, which might not be separable. As shown in \cite{CEMMSE} for the time-frequency case this leads to a minimal performance loss, but with significantly lower complexity. 
In many cases the covariance is unknown, and one would need to generate the interpolation martrices based on some model for the covariance, whose parameters would also then have to be estimated.
It is important to mention that the same interpolation matrix $\boldsymbol{A}_f$ is used for all \ac*{OFDM} symbols containing reference symbols. This means that if the reference signal pattern in 
these symbols is not the same, we need to apply different matrices for each symbol. But fortunately this assumption holds true for the chosen 3GPP NR \ac*{OFDM} type 1 \ac*{DMRS}.

For the example in this work we assume the channel remains constant in one subframe of 14 \ac*{OFDM} symbols (Doppler spread equal to zero). 
Therefore, the time interpolation matrix $\boldsymbol{A}_t$ consists only of a averaging among the \ac*{OFDM} symbols that contain reference symbols. 

The frequency interpolation and smoothing matrix $\boldsymbol{A}_f$ is based on the \ac*{MMSE} solution described in \cite{CEMMSE}. 
To generate these matrices based on this method, we need to generate the auto and cross correlation of the channel $\boldsymbol{R}_{\boldsymbol{h}_d\boldsymbol{h}_d}$ and $\boldsymbol{R}_{\boldsymbol{h}_d\boldsymbol{h}_\ell}$.
The symbols $\boldsymbol{h}_{d}$ and $\boldsymbol{h}_\ell$ are the vector of the channel at the position of the reference signals and the channel on all subcarriers on \ac*{OFDM} symbol $\ell$.
In the case that we know all these matrices, the interpolation matrix is defined as
\begin{equation}
	\boldsymbol{A}_f = \boldsymbol{R}_{\boldsymbol{h}_d\boldsymbol{h}_\ell} \left( \boldsymbol{R}_{\boldsymbol{h}_d\boldsymbol{h}_d} + \boldsymbol{R}_{\boldsymbol{\eta}\boldsymbol{\eta}}\right)^{-1},
\end{equation}
where $\boldsymbol{R}_{\boldsymbol{\eta}\boldsymbol{\eta}}$ is the noise covariance matrix. In a practical system we could easily have hundreds of subcarriers, thus the complexity of the matrix inversion is extreme.
Even if the values of the inversion are precomputed the following matrix vector multiplication also has a high complexity. We therefore limit the length of reference symbols considered for the interpolation to $K_C$. 
This does decrease the complexity and has only a minor impact on the performance, since the channel that have a large distance in terms of subcarriers are close to uncorrelated. 

In addition for a practical system is in many cases not possible to directly observe and estimate the covariance matrix of the channel, especially at the position with no reference signals present. 
Consequently, we use a model for the generation of the covariance matrices. For many real world scenarios the channel path arriving later at the receiver propagate through a longer path, thus leading to lower energy at the receiver. 
This can be well approximated by a exponential \ac*{PDP}.

Since all the elements of $\boldsymbol{R}_{\boldsymbol{h}_d\boldsymbol{h}_\ell}$ and $\boldsymbol{R}_{\boldsymbol{h}_d\boldsymbol{h}_d}$ represent the cross correlation of different elements of the channel.
Thus, all elements of these matrices are defined by the cross correlation between the channels on subcarrier $i$ and $j$
\begin{equation}
	\mathbb{E}[h_i h^*_j] = \frac{1}{1 - j 2 \pi \tau_{\text{RMS}} \Delta_f d(i, j)},
\end{equation}
where $\tau_{\text{RMS}}$ is the \ac*{RMS} delay spread and $d(i, j)$ the distance of the $i$th to the $j$th subcarrier.

\section{MIMO Detection Algorithm}
In this section we show how the sequential \ac*{DCD} with bound is derived from a relaxation of the \ac*{ML} MIMO equalization. 
The classical problem of \ac*{ML} detection can be formulated as
\begin{equation}
 \hat{\boldsymbol{x}} = \argmin\limits_{x_n \in \mathbb{X}} \left\vert\left\vert \boldsymbol{y} - \boldsymbol{H}\boldsymbol{x}\right\vert\right\vert^2_2,
\end{equation}
where $\mathbb{X}$ is the set containing all possible transmit symbols. 
The symbols $\boldsymbol{x}$, $\boldsymbol{H}$, $\boldsymbol{y}$ and $\hat{\boldsymbol{x}}$ represent the transmit symbol, the channel, the receive symbol and the symbol after the detection of the system.
The complexity of this discrete optimization problem grows exponentially with the dimensions of $\boldsymbol{x}$ and the size of the set $\mathbb{X}$.
Thus, for higher number of spatial streams as envisioned for massive \ac*{MIMO} it is not feasible to solve this problem. 

Fortunately, as the number of receive antennas grows large with respects the number of simultaneously served users, the \ac*{MMSE} solution to the relaxed optimization problem
\begin{equation}
 \hat{\boldsymbol{x}} = \argmin \mathbb{E}\left[\left\vert\left\vert \boldsymbol{y} - \boldsymbol{H}\boldsymbol{x}\right\vert\right\vert^2_2\right],
\end{equation}
approaches the performance of the \ac*{ML} detection \cite{Yang2015}.
Unfortunately, the close form solution to this problem requires knowledge about the noise covariance matrix. For a system with a large number of antennas this is hard to attain. 

Therefore, we choose to relax the ML detection problem in a ways that reduces the complexity, but not making any assumptions on the noise statistics
\begin{equation}
 \hat{\boldsymbol{x}} = \argmin\limits_{\Re(x_n) \in[-B, B], ~\Im(x_n) \in[-B, B]} \left\vert\left\vert \boldsymbol{y} - \boldsymbol{H}\boldsymbol{x}\right\vert\right\vert^2_2.
\end{equation}
The variable $B$ forces the real and imaginary part of each element of the vector $\hat{\boldsymbol{x}}$ to be in the range from $-B$ to $B$. 
In the following paragraphs we will show how to solve this optimization problem efficiently and that we do not need to make any assumption on the noise statistics.

\begin{figure}
\vspace*{-0.1cm}
\begin{algorithm}[H]
	\caption{Sequential DCD with bound}
	\begin{algorithmic}
		\State \textbf{Require:} $\boldsymbol{A}$, $\boldsymbol{b}$, $N$, $H$, $B$, $N_u$, $M_b$
		\State \textbf{Initialization:} $\boldsymbol{x} \gets \boldsymbol{0}$, $\boldsymbol{r} \gets \boldsymbol{b}$, $\alpha \gets H$, $m \gets 0$, \\ UpdateFlag $\gets$ false, $k \gets 0$
		\While{$m < M_b$}
			\For{$n \in \{1, \cdots, N\}$}
				\If{$\alpha/2 [\boldsymbol{A}]_{n,n} < \vert [\boldsymbol{r}]_n\vert$}
					\State $t \gets [\boldsymbol{x}]_n + \text{sign}([\boldsymbol{r}]_n)\alpha $
					\If{$t \le B$}
						\State $[\boldsymbol{x}]_n \gets t$
						\State $\boldsymbol{r} \gets \boldsymbol{r} -  \text{sign}([\boldsymbol{r}]_n)\alpha \boldsymbol{a}_n$
						\State UpdateFlag $\gets$ true, $k \gets k + 1$
					\EndIf
				\EndIf
			\EndFor
			\If{$k \ge N_u$}
				\State \Return{$\boldsymbol{x}$, $\boldsymbol{r}$}
			\EndIf
			\If{UpdateFlag}
				\State UpdateFlag $\gets$ false
			\Else
				\State $m \gets m + 1$, $\alpha \gets \alpha / 2$
			\EndIf
		\EndWhile
		\State \Return{$\boldsymbol{x}$, $\boldsymbol{r}$}
  \end{algorithmic}
  \label{alg:DCDBound}
\end{algorithm}
\vspace*{-0.3cm}
\end{figure}

This problem can be reformulated into solving the following linear system of equations
\begin{equation}
\boldsymbol{H}^H\boldsymbol{H}\boldsymbol{x} = \boldsymbol{H}^H\boldsymbol{y}.
\end{equation}
Thus, we can utilize a coordinate descend based method to solve this problem.
To reduce the complexity we selected the step-size to be of the form $2^{-l}$, where $l$ is an integer. 
This has the advantage that all multiplications with this number can be implemented as bit shifts. 
These algorithms are called \ac*{DCD} as described in \cite{Quan2007, Quan2009} for multi user detection in a \ac*{CDMA} system and is shown in Algorithm \ref{alg:DCDBound}.
The parameters $H$, $B$, $N_u$ and $M_b$ are the maximum step-size, the upper bound of detected symbols, the maximum number of updates and the maximum number of step-size divisions by 2 of the algorithm. 
The parameter $B$ should be chosen in a way to just accommodate the \ac*{QAM} constellation in the scaling before the data detection.
For the case that the constellation is not bounded this parameters can be set to a reasonable large value or even infinity.
A \ac*{DFT-s-OFDM} system is a example for a constellation that is not bounded when the data detection is calculated in the frequency domain. 
The value of $H$ should be of the form $2^{-l}$, where $l$ is an integer. 
Since it is not useful to start with a step-size that is larger than the final bound, $H$ should also be smaller than $B$.
A good way to choose $H$ would be $H = 2^{\lfloor \log_2(B)\rfloor}$.
This ensures that we start with the maximum possible step-size to enable fast convergence.
The symbols $\boldsymbol{A}$, $\boldsymbol{b}$ and $N$ define the linear systems of equations
\begin{equation}
\boldsymbol{A}\boldsymbol{x} = \boldsymbol{b},
\end{equation}
and the size of the vector $\boldsymbol{b}$. 
The vectors $\boldsymbol{x}$ and $\boldsymbol{r}$ represent the resulting vector and the residual error, which is updated in every step. 
Since algorithm \ref{alg:DCDBound} solves only real linear systems of equations $\boldsymbol{A}$ and $\boldsymbol{b}$ 
are related to $\boldsymbol{H}$ and $\boldsymbol{y}$ in the following way
\begin{equation}
\boldsymbol{A} = 
 \begin{bmatrix}
  \Re(\boldsymbol{H}^H\boldsymbol{H}) & -\Im(\boldsymbol{H}^H\boldsymbol{H}) \\
  \Im(\boldsymbol{H}^H\boldsymbol{H}) & \Re(\boldsymbol{H}^H\boldsymbol{H}) \\
 \end{bmatrix},
\boldsymbol{b} =
  \begin{bmatrix}
  \Re(\boldsymbol{H}^H\boldsymbol{y}) \\
  \Im(\boldsymbol{H}^H\boldsymbol{y}) \\
 \end{bmatrix}.
\end{equation}
Since we solve the equivalent real linear system of equations the value of $N$ is double the number of users/spatial streams to be detected. The resulting value of $\boldsymbol{x}$ is also going to be split between real and imaginary part in the same way as $\boldsymbol{b}$. A simple recombination of $\boldsymbol{x}$ leads to detected symbol.

\section{Simulation Results}

In this section we compare the performance of sequential \ac*{DCD} to \ac*{MMSE} equalization. 
Since we consider system with low resolution \acp*{ADC} and channel estimation error, the noise is in general not white. 
We also tested additional noise whitening for both system, but since no performance gain was observed at an addition computational cost, we only show results without noise whitening. 
The simulation parameter in Table \ref{tab:SimConfig} shows the most important simulation parameters.
 It is also important to mention that the bounds for the sequential \ac*{DCD} equalization tightly encloses the QAM constellation.
\begin{table}
	\renewcommand{\arraystretch}{1.3}
	\caption{Simulation Parameters. }
	\label{tab:SimConfig}
	\centering
		\begin{tabular}{|p{4cm}|p{4cm}|}
			\hline
			Parameter & Value \\ \hline \hline
			Reference Signal & 3GPP NR OFDM type 1 DMRS \\ \hline
			Channel Estimation & 2x1D MMSE and ideal \\ \hline
			Number of Users & 8 \\ \hline
			Number of receive antennas & 64 \\ \hline
			Channel model & Exponential PDP (no Doppler spreach) \\ \hline
			SNR definition & Average per user per antenna SNR \\ \hline
			Channel code & LTE turbo code with rate 0.9 \\ \hline
			MIMO detection algorithms & MMSE and DCD-Bound \\ \hline
			ADC resolution & 2 bit \\ \hline
			Modulation format & 16 QAM \\ \hline
		\end{tabular}
\end{table}
\subsection{Performance Results}

\begin{figure}
\begin{center}
\begin{tikzpicture}
    \begin{axis}[width=0.95*\columnwidth, height=6cm, ylabel={coded BER}, grid,  legend cell align=left, ymode=log,
                  		xlabel={SNR [dB]}, xmin =-20, xmax=0, ymin =1e-3, ymax = 1, legend pos=south west, legend cell align=left, legend columns=1]

		\addplot[ thick, red, mark=*, mark options={solid}] table[x = x, y=y]{SimulationResults/PlotData/CodedBERMMSEChest.txt};
		\addlegendentry{MMSE Chest}
		\addplot[ thick, blue, dashed, mark=*, mark options={solid}] table[x = x, y=y]{SimulationResults/PlotData/CodedBERDCDBoundChest.txt};
		\addlegendentry{DCD Chest}
		\addplot[ thick, red] table[x = x, y=y]{SimulationResults/PlotData/CodedBERMMSEIdeal.txt};
		\addlegendentry{MMSE Ideal}
		\addplot[ thick, blue, dashed] table[x = x, y=y]{SimulationResults/PlotData/CodedBERDCDBoundIdeal.txt};
		\addlegendentry{DCD Ideal}

     \end{axis}
\end{tikzpicture}
\end{center}
\vspace*{-0.5cm}
\caption{Coded BER with LTE turbo code rate = 0.9.}
\label{fig:CodedBER}
\end{figure}
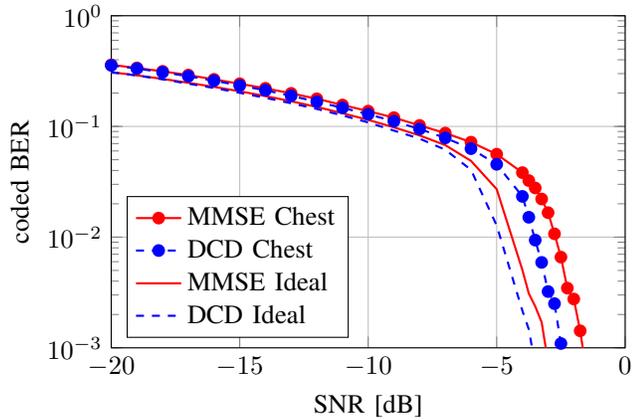

The uncoded and coded \ac*{BER} results are shown in Fig. \ref{fig:UncodedBER} and \ref{fig:CodedBER}. 
Due to the frequency selective channel and the 2-bit resolution \acp*{ADC}, the uncoded \ac*{BER} does exhibits a error floor. 
It is also obvious that error floor of the \ac*{DCD} is lower then the one of the \ac*{MMSE} algorithm w/o ideal channel estimation.
As we can see from the zoomed in part around $10^{-2}$ \ac*{BER}, the performance of \ac*{DCD} is slightly more robust to channel estimation errors. 
Fig. \ref{fig:CodedBER} shows that these results translate well to a system with channel coding. 
\begin{figure}
\begin{center}
\begin{tikzpicture}
    \begin{axis}[width=0.95*\columnwidth, height=6cm, ylabel={uncoded BER}, grid,  legend cell align=left, ymode=log, name=mainplot,
                  		xlabel={SNR [dB]}, xmin =-20, xmax=10, ymin =5e-4, ymax = 1, legend pos=south west, legend cell align=left, legend columns=1]

		\addplot[ thick, red, mark=*, mark options={solid}, mark repeat*=5] table[x = x, y=y]{SimulationResults/PlotData/UncodedBERMMSEChest.txt};
		\addlegendentry{MMSE Chest}
		\addplot[ thick, blue, dashed, mark=*, mark options={solid}, mark repeat*=5] table[x = x, y=y]{SimulationResults/PlotData/UncodedBERDCDBoundChest.txt};
		\addlegendentry{DCD Chest}
		\addplot[ thick, red] table[x = x, y=y]{SimulationResults/PlotData/UncodedBERMMSEIdeal.txt};
		\addlegendentry{MMSE Ideal}
		\addplot[ thick, blue, dashed] table[x = x, y=y]{SimulationResults/PlotData/UncodedBERDCDBoundIdeal.txt};
		\addlegendentry{DCD Ideal}

     \end{axis}
     \begin{axis}[at={(6.835cm,4.425cm)}, anchor={outer north east}, width=3cm, height=2cm,  legend cell align=left, ymode=log, name=inset, 
     				xtick={-3, -2,  -1,  0}, minor xtick={-3, -3, -2.5, -1.5, -0.5, 0.5}, grid=both, ticklabel style = {font=\tiny}, 
     				xticklabel style = {xshift=0.05cm, yshift=0.08cm, font=\tiny},
                  		xmin =-3, xmax=0.5, ymin =9e-3, ymax = 1.1e-2, scale only axis, axis background/.style={fill=white}]

		\addplot[ thick, red, mark=*, mark options={solid}] table[x = x, y=y]{SimulationResults/PlotData/UncodedBERMMSEChest.txt};
		\addplot[ thick, blue, dashed, mark=*, mark options={solid}] table[x = x, y=y]{SimulationResults/PlotData/UncodedBERDCDBoundChest.txt};
		\addplot[ thick, red] table[x = x, y=y]{SimulationResults/PlotData/UncodedBERMMSEIdeal.txt};
		\addplot[ thick, blue, dashed] table[x = x, y=y]{SimulationResults/PlotData/UncodedBERDCDBoundIdeal.txt};
     \end{axis}
\end{tikzpicture}
\end{center}
\vspace*{-0.5cm}
\caption{Uncoded BER results}
\label{fig:UncodedBER}
\end{figure}
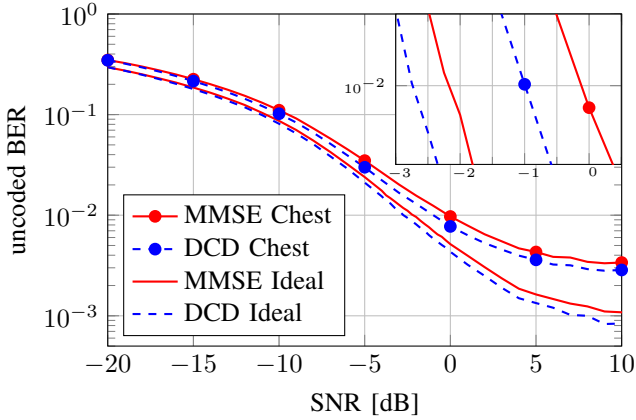

\subsection{Computational Complexity Results}
\begin{figure}
\centering
\begin{tikzpicture}
\begin{axis}[%
width=0.95*\columnwidth,
height=6cm,
xmin=1000,
xmax=3000,
ymin=0,
ymax=60000,
axis background/.style={fill=white},
xtick={1000, 1500, 2000, 2500, 3000},
xticklabels={1000, 1500, 2000, 2500, 3000},
xmajorgrids,
ymajorgrids
]
\addplot[area legend, table/row sep=crcr, patch, patch type=rectangle, shader=flat corner, color=magenta!30, draw=black, forget plot, patch table with point meta={%
1	2	3	4	1\\
6	7	8	9	1\\
11	12	13	14	1\\
16	17	18	19	1\\
21	22	23	24	1\\
26	27	28	29	1\\
31	32	33	34	1\\
36	37	38	39	1\\
41	42	43	44	1\\
46	47	48	49	1\\
51	52	53	54	1\\
56	57	58	59	1\\
61	62	63	64	1\\
66	67	68	69	1\\
71	72	73	74	1\\
76	77	78	79	1\\
81	82	83	84	1\\
86	87	88	89	1\\
91	92	93	94	1\\
96	97	98	99	1\\
101	102	103	104	1\\
106	107	108	109	1\\
111	112	113	114	1\\
116	117	118	119	1\\
121	122	123	124	1\\
126	127	128	129	1\\
131	132	133	134	1\\
136	137	138	139	1\\
141	142	143	144	1\\
146	147	148	149	1\\
}]
table[row sep=crcr] {%
x	y\\
1040	0\\
1040	0\\
1040	1\\
1106.66666666667	1\\
1106.66666666667	0\\
1106.66666666667	0\\
1106.66666666667	0\\
1106.66666666667	3\\
1173.33333333333	3\\
1173.33333333333	0\\
1173.33333333333	0\\
1173.33333333333	0\\
1173.33333333333	21\\
1240	21\\
1240	0\\
1240	0\\
1240	0\\
1240	93\\
1306.66666666667	93\\
1306.66666666667	0\\
1306.66666666667	0\\
1306.66666666667	0\\
1306.66666666667	310\\
1373.33333333333	310\\
1373.33333333333	0\\
1373.33333333333	0\\
1373.33333333333	0\\
1373.33333333333	1386\\
1440	1386\\
1440	0\\
1440	0\\
1440	0\\
1440	3250\\
1506.66666666667	3250\\
1506.66666666667	0\\
1506.66666666667	0\\
1506.66666666667	0\\
1506.66666666667	6869\\
1573.33333333333	6869\\
1573.33333333333	0\\
1573.33333333333	0\\
1573.33333333333	0\\
1573.33333333333	13125\\
1640	13125\\
1640	0\\
1640	0\\
1640	0\\
1640	21638\\
1706.66666666667	21638\\
1706.66666666667	0\\
1706.66666666667	0\\
1706.66666666667	0\\
1706.66666666667	32225\\
1773.33333333333	32225\\
1773.33333333333	0\\
1773.33333333333	0\\
1773.33333333333	0\\
1773.33333333333	55696\\
1840	55696\\
1840	0\\
1840	0\\
1840	0\\
1840	53980\\
1906.66666666667	53980\\
1906.66666666667	0\\
1906.66666666667	0\\
1906.66666666667	0\\
1906.66666666667	57899\\
1973.33333333333	57899\\
1973.33333333333	0\\
1973.33333333333	0\\
1973.33333333333	0\\
1973.33333333333	56878\\
2040	56878\\
2040	0\\
2040	0\\
2040	0\\
2040	50705\\
2106.66666666667	50705\\
2106.66666666667	0\\
2106.66666666667	0\\
2106.66666666667	0\\
2106.66666666667	41385\\
2173.33333333333	41385\\
2173.33333333333	0\\
2173.33333333333	0\\
2173.33333333333	0\\
2173.33333333333	37838\\
2240	37838\\
2240	0\\
2240	0\\
2240	0\\
2240	19136\\
2306.66666666667	19136\\
2306.66666666667	0\\
2306.66666666667	0\\
2306.66666666667	0\\
2306.66666666667	12032\\
2373.33333333333	12032\\
2373.33333333333	0\\
2373.33333333333	0\\
2373.33333333333	0\\
2373.33333333333	7196\\
2440	7196\\
2440	0\\
2440	0\\
2440	0\\
2440	3740\\
2506.66666666667	3740\\
2506.66666666667	0\\
2506.66666666667	0\\
2506.66666666667	0\\
2506.66666666667	1897\\
2573.33333333333	1897\\
2573.33333333333	0\\
2573.33333333333	0\\
2573.33333333333	0\\
2573.33333333333	1009\\
2640	1009\\
2640	0\\
2640	0\\
2640	0\\
2640	293\\
2706.66666666667	293\\
2706.66666666667	0\\
2706.66666666667	0\\
2706.66666666667	0\\
2706.66666666667	139\\
2773.33333333333	139\\
2773.33333333333	0\\
2773.33333333333	0\\
2773.33333333333	0\\
2773.33333333333	42\\
2840	42\\
2840	0\\
2840	0\\
2840	0\\
2840	8\\
2906.66666666667	8\\
2906.66666666667	0\\
2906.66666666667	0\\
2906.66666666667	0\\
2906.66666666667	3\\
2973.33333333333	3\\
2973.33333333333	0\\
2973.33333333333	0\\
2973.33333333333	0\\
2973.33333333333	3\\
3040	3\\
3040	0\\
3040	0\\
};
\end{axis}
\end{tikzpicture}%
\vspace*{-0.2cm}
\caption{Histogram of the additions require to converge.}
\label{fig:ComplexityHist}
\end{figure}
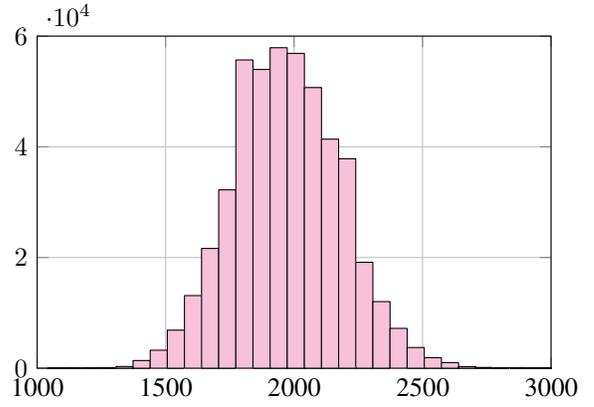
\begin{table}
	\renewcommand{\arraystretch}{1.3}
	\caption{Complexity per operation.}
	\label{tab:OperationComplex}
	\centering
		\begin{tabular}{|p{3cm}|p{1.2cm}|p{1.2cm}|p{1.2cm}|}
			\hline
			operation & real additions & real multiplications & logic operations \\ \hline \hline
			$\boldsymbol{H}^H\boldsymbol{H}$ & 8128 & 8192 & 19038400 \\ \hline
			$\boldsymbol{h}^H\boldsymbol{y}$ & 2032 & 2048 & 5521600 \\ \hline
			$\boldsymbol{H}^H\boldsymbol{H} + \boldsymbol{I}$ & 16 & 0 & 2000\\ \hline
			$(\boldsymbol{H}^H\boldsymbol{H} + \boldsymbol{I})^{-1}$ & 1700 & 1900 & 4392500 \\ \hline
			$(\boldsymbol{H}^H\boldsymbol{H} + \boldsymbol{I})^{-1}\boldsymbol{h}^H\boldsymbol{y}$ & 240 & 256 & 593200 \\ \hline
			Sequential \ac*{DCD} with bound & 2000 & 0 & 250000 \\ \hline
		\end{tabular}
\end{table}
\begin{table}
	\renewcommand{\arraystretch}{1.3}
	\caption{Complexity results.}
	\label{tab:CplxResults}
	\centering
		\begin{tabular}{|p{3.5cm}|p{1.6cm}|p{1.6cm}|}
			\hline
			detection algorithm & scenario 1 & scenario 2 \\ \hline \hline
			\ac*{MMSE} & 29547700 & 109040100 \\ \hline
			Sequential \ac*{DCD} with bound & 24810000 & 99840800 \\ \hline
		\end{tabular}
\end{table}

Since the \ac*{DCD} algorithm has no multiplications we need to compare the complexity to \ac*{MMSE} by mapping additions and multiplications to logic operations. 
The work in \cite{IPGATES2017, Bilavarn2006} offer a mapping of real additions and multiplications to logic gates. Assuming 18 bit signed fixed point calculation an
addition and a multiplication can be implemented using 125 and 2200 NAND gates with two inputs, respectively. 
To compare the different algorithms, we compare this number of logic gates as the number of logic operation required to calculate the result. 

As the sequential \ac*{DCD} algorithm is an iterative procedure we need to investigate the converges of it. The histogram in Fig. \ref{fig:ComplexityHist} shows the number of addition used to process one symbol vector for the simulation parameters presented in the preceding paragraph. The comparisons are implemented as a subtraction followed by checking if the sign bit is set or not. Therefore, they are counted to have equal complexity compared to an addition. The number of comparisons is low compared to additions used for updating the residual vector $\boldsymbol{r}$. The average complexity is 1972 real additions. For simplicity we use 2000 for the following analysis. 

From the Table \ref{tab:OperationComplex} it is easy to see that even just the multiplication with the already inverted matrix is more complex the solving the linear system of equations with the Sequential \ac*{DCD} algorithm with bound.
It is also obvious that the complexity is dominated by the computation of the Gram matrix ($\boldsymbol{H}^H\boldsymbol{H}$). It is important to mention that for the calculation of the Gram matrix, we already exploited the symmetry of the resulting matrix to minimize the necessary multiplications and additions. 
All other computation steps have a much lower complexity. In this investigation, we neglected the necessary complexity for normalization of the signal power and the additional complexity for making the \ac*{MMSE} equalizer unbiased. 

To Compare the \ac*{MMSE} equalizer to the algorithm developed here, we compare two scenarios. In the first scenario, the matrix computed to generate the \ac*{MMSE} result is calculated separately for each sub-carrier (scenario 1). In the second case we assume that the matrix can be reused to detect the symbol in 14 consecutive \ac*{OFDM} symbols on the same sub-carrier (scenario 2). 
There are few common operations to both systems and we assume that this intermediate calculations can be stored and reused for scenario 2. 
The complexity for both algorithms are shown in Table \ref{tab:CplxResults}. The overall computational complexity of our approach compared to \ac*{MMSE} is reduced while at the same time the performance is improved. In the first scenario the improvement is about 16\% in the second it is in the range of 10\%. 

\section{Conclusion}
Our investigation showed that a bounded \ac*{DCD} MIMO equalization algorithm does outperform a \ac*{MMSE} based equalization. In addition, sequential \ac*{DCD} has a lower computational complexity. 
But it is important to mention that in contrast to many other papers considering the complexity for massive \ac*{MIMO} we showed that the 
complexity is dominated by the computation of the gram matrix and not the matrix inversion.
This evaluation shows that it is possible to achieve high data rates with digital beamforming \ac*{mmWave} system with low resolution \acp*{ADC} by considering low complexity algorithms.

\section*{Acknowledgment}
This work has been performed in the framework of the Horizon 2020 project ONE5G (ICT-760809) receiving funds from the European Union. The authors would like to acknowledge the contributions of their colleagues in the project, although the views expressed in this contribution are those of the authors and do not necessarily represent the project.


\bibliographystyle{IEEEtran}
\bibliography{IEEEabrv,../../../bibliography/bibKilian}

\end{document}